\documentstyle[preprint,aps,tighten,epsfig]{revtex}
\begin{document}

\preprint{
 \parbox{1.5in}{\leftline{JLAB-THY-97-08}
                \leftline{WM-97-106} 
                \leftline{CFNUL-97-02}  }  } 
\title{ Effects of Relativistic Dynamics in 
$pp \longrightarrow pp \pi^0$\ near Threshold}

\author{J.\ Adam, Jr.\ $^{1,2}$,  Alfred Stadler $^3$,
        M.\ T.\ Pe\~{n}a$^{3,4}$ , Franz Gross$^{1,5}$ }
\address{
$^1$Jefferson Lab, 
     12000 Jefferson Avenue, Newport News, VA 23606\\
$^2$Institute of Nuclear Physics, \v{R}e\v{z} n.\ Prague, 
     CZ-25068, Czech Republic\\
$^3$Centro de F\'{\i}sica Nuclear da Universidade de Lisboa, 
     1699 Lisboa Codex, Portugal\\ 
$^4$Departamento de F\'{\i}sica, Instituto Superior T\'ecnico, 
     1096 Lisboa Codex, Portugal\\ 
$^5$Department of Physics, College of William and Mary,
    Williamsburg, Virginia 23185 }

\date{\today}
\maketitle

\begin{abstract}
The cross-section for threshold $\pi^0$\ production in proton-proton
collisions is evaluated in the framework of the covariant spectator description.
The negative energy intermediate states are included 
non-perturbatively and seen to yield a considerably smaller contribution, 
when compared to perturbative treatments. A family of OBE-models with 
different off-shell scalar coupling is considered.  
\end{abstract}

\narrowtext

\section{Introduction}

The reaction $pp \longrightarrow pp \pi^0$\ near threshold is very sensitive 
to the description of the $NN$-interaction and to $\pi^0$\ production mechanisms.
 Due to the underlying chiral symmetry both
the single-nucleon term and the pion-rescattering contribution are
suppressed near threshold \cite{Kol96a,Kol96b,Park}. The transition amplitude
then results from a delicate interference between these  
and various additional contributions of shorter range  
\cite{Kol96b,KR66,MS91,Nis92,Hor93,Lee93,Oset,Hanh}.
The relevance and relative importance of various reaction mechanisms, in
particular the short range ones, has not yet been firmly established. 
A treatment of these mechanisms consistent with the $NN$ interactions 
employed in the description of the  final and initial state
is needed to clarify the large model dependence. 
Moreover, the magnitude of the transferred  hadronic
momentum, which is typically $p/m \sim \sqrt{\mu / m } \sim 0.4$, 
calls for a relativistic approach.

In this letter we re-examine the single-nucleon term using the covariant
spectator  description \cite{spec,GVH,StG} for both  
the production mechanism and for the  initial
and final state $pp$\ interaction. 
The pion emission from a single proton is described by the Feynman
diagrams of Figure 1.  The OBE spectator models \cite{GVH} 
allow a direct covariant evaluation of diagrams (a)-(c). Dynamically,
these are the only model-independent contributions. There might be
additional reaction mechanisms, in particular involving  pion
rescattering, such as diagram (d). They have not yet been included in
our calculations.
The intermediate state nucleons in  diagrams (a)-(c) can propagate with 
positive and negative energy. We will split the contribution of
these diagrams into amplitudes  $a^{\pm}, b^{\pm} $\ and $c^{\pm,\pm}$\
with the superscript indicating the sign of the energy of the off-shell
nucleon after and/or before the pion emission.   

Our approach differs crucially  from earlier ones in two aspects. 
Firstly, it includes nonperturbatively diagrams with intermediate 
negative-energy nucleons. In perturbative approaches these contributions
are simulated by the inclusion of  effective meson-exchange operators
acting in two-nucleon space (their contributions are often  called
 ``Z-diagrams''). More specifically, 
one takes the amplitudes $a^-$\ and $b^-$\  (with 
the nucleons in the initial and final state on-shell) to
the lowest order in $v/c \sim p/m$, replacing  the $NN$ interaction 
by a single one-meson exchange, usually taking into account only
the most important meson exchanges. Written in the space of  
two-component Pauli spinors these amplitudes are identified with 
 effective pion-production meson-exchange operators  and their 
expectation values are then calculated  with conventional nuclear 
wave functions. These transition amplitudes simulate those 
contributions of the diagrams  $a^-, b^-, c^{+-}$\ and $c^{-+}$\ 
for which the transition to negative-energy nucleons occurs
only once.
 
In previous perturbative calculations
\cite{Kol96b,Park,KR66,MS91,Nis92,Hor93,Lee93,Oset,Hanh}  the most 
important contributions to the Z-diagram operator were found 
to be the exchanges of $\sigma$, $\omega$, and $\pi$\ mesons. It has been 
shown \cite{Hor93} that the pion Z-graph yields a much
smaller contribution than the $\sigma$\ and $\omega$\ diagrams.
It is often considered together with other model dependent
off-shell pion-rescattering processes
\cite{Kol96b,Oset,Hanh} . In $v/c$-expansion calculations,
the $\sigma$\ and $\omega$\ Z-graphs were found to dominate
the short range reaction mechanisms and provide contributions
essential for the explanation of the data. In contrast,
in the present covariant calculations we obtained significantly 
smaller contributions from terms with  negative-energy nucleons.   
 
Secondly, the $NN$-models considered recently \cite{StG} include non-minimal 
dynamical effects produced by off-shell scalar-nucleon coupling, scaled  
by  a parameter $\nu$:
\begin{equation}
 \Gamma_{s} (p^{\, \prime}, p) = g_s \left\{ 1 + \frac{\nu_s}{2 m} \, \left[
  \not{\! p}^{\, \prime} + \not{\! p} - 2 m \, \right] \right\} \, ,
\end{equation}
with $\nu_{\sigma}= -0.75\, \nu$, and $\nu_{\delta}=  2.6 \, \nu$.  
Both $NN$-observables and the triton binding energy 
appear to be rather sensitive 
to this off-shell extension, hence it is very  interesting to find out if 
it also works favorably in the considered  reaction. Taking only 
diagrams (a) and (b)  for $\sigma$\ and $\omega$\ exchange 
in Born approximation, we, indeed, have  
found a rather strong dependence on the parameter $\nu$,
mainly due to the cancellation between  $\sigma$\ and $\omega$\ terms. 
However, the variation is much weaker when the full interaction is
included.   

\section{Theoretical description}
 
Near threshold the reaction is dominated by transition between initial 
$^3P_0$\ and final $^1S_0$\ \,  $pp$\   partial waves; 
only these are considered in our calculations.  The total cross section 
is given \cite{BD} in terms of the 
covariant transition amplitude $\cal{M}$:
\begin{eqnarray}
 \sigma [\mu b] &=& \frac{10^4 (\hbar c)^2}{(2 \pi)^5} \,  
  \frac{m^2}{4 p E} \, \,   \overline{ |{\cal M}|^2 }  
  \, \, \rho (T_{lab} )  \, , 
\label{cross}  \\
 \overline{ |{\cal M}|^2 } &=& \frac{1}{4} 
  \sum_{\lambda_1^{\, \prime}, \lambda_1, \lambda_2^{\, \prime}, \lambda_2}  
 {\cal M}_{\lambda_1^{\, \prime} \lambda_1, \lambda_2^{\, \prime} \lambda_2}     
 \, \, 
{\cal M}^*_{\lambda_1^{\, \prime} \lambda_1, \lambda_2^{\, \prime} \lambda_2}       
  \,  ,
\label{amp} 
\end{eqnarray} 
where $m$\ is the nucleon mass; $p, E= \sqrt{p^2 + m^2}$\ are the relative
momentum and energy in the initial state; $\rho(T_{lab})$\ is the phase
space density, $T_{lab}$\ is the kinetic energy of the incoming proton.  
The amplitude $\cal{M}$\ and the phase space density $\rho$\ have 
dimensions MeV$^{-3}$, and MeV$^{4}$, respectively.
The helicity transition amplitude 
${\cal M}_{\lambda_1^{\, \prime} \lambda_1, \lambda_2^{\, \prime} \lambda_2} $
is calculated from the Feynman rules as defined in \cite{BD}. For the partial
waves considered, there is only one independent helicity amplitude,
${\cal M}^{J=0}_{++,++} = i \exp (i \, \delta_3 (E))\, {\cal M} $, 
with $\delta_3 (E)$\ being the $^3P_0$\ phase shift.  

The $NN$-interactions \cite{GVH} used in our calculations
were fitted to $np$-data and the effect of the Coulomb
interaction is not included. It would suppress the cross section at 
threshold by about $ 40 \% $. It has been shown   \cite{MS91}, 
that using a constant amplitude in (\ref{cross}) 
results in a roughly  quadratic dependence
of $\sigma$\ on $T_{lab}$, clearly at variance with the experimental
data (Fig. \ref{fig2}).
 
The most pronounced energy dependence of the total transition amplitude
comes from the energy dependence of the final state $^1S_0$\ interaction.
We deal with it in an approximate way similar to that employed in \cite{Meyer}:
we assume that the energy dependence of the final state scattering matrix 
is well approximated by the energy dependence of the
corresponding on-shell $NN$ scattering matrix $M(p)= M(p,p)$, 
\begin{equation}
 M ( p^{\, \prime}, p )= M (0,p) \frac{ M (p^{\, \prime}, p) }{M (0, p)}
 \approx M (0, p) \frac{M(p^{\, \prime})}{ M(0)}  \, .
\label{mmat}
\end{equation}
Expressing $M(p^{\, \prime})$\ in terms of the $^1S_0$
phase-shift $\delta_1$ and $M(0)$\ in terms of the 
singlet scattering length $a_s$, one obtains
\begin{equation}
 M ( p^{\, \prime}, p ) \approx - \exp (i \delta_1 (E^{\, \prime}) ) 
 \frac{m}{E^{\, \prime}}
 \frac{ \sin (\delta_1 (E^{\, \prime}))}{(a_s p^{\, \prime}\, )}
  M ( 0, p) \, ,
\end{equation}
with $ p^{\, \prime}, E^{\, \prime}$\ being the relative momentum
and energy of the final $pp$\ state.
Accordingly, the square of the transition amplitude
contains the energy-dependent factor
\begin{equation}
 \xi (p^{\, \prime}\, ) = (\frac{m}{E^{\, \prime}})^2
 \frac{ \sin^2 (\delta_1 (E^{\, \prime}))}{(a_s p^{\, \prime}\, )^2} \, .
\end{equation}
We introduce it into (\ref{cross}) by replacing 
\begin{eqnarray}
 \rho (T_{lab}) & \longrightarrow  &\rho (\xi, T_{lab}) =
 \int  
 \frac{m}{E_1^{\, \prime}}  d \vec{p}_1^{\, \prime}   \, 
 \frac{m}{E_2^{\, \prime}}  d \vec{p}_2^{\, \prime}   \, 
 \frac{d \vec{q}\, }{2 \omega_q}  \nonumber\\
 &&  \xi (p^{\, \prime}\, ) \, \,
  \delta^{(4)} ( p_1 + p_2 - p_1^{\, \prime}- p_1^{\, \prime}- q) \, .
\label{phase}
\end{eqnarray}

The difference between 
$\rho (1,T_{lab})$\ and $\rho (\xi,T_{lab})$\ is illustrated in 
Fig. \ref{fig2}. The approximation  introduced in (\ref{mmat})  
implies  for the Heitler $R$-matrix 
$ R(p^{\, \prime}, p)/R(p^{\, \prime}) \approx
R(0, p)/R(0)$.  Figure 3 demonstrates  
the quality of this assumption about the energy 
dependence of the off-shell $R$-matrices.
The curves corresponding to different energies would coincide in case
the factorization describing the energy dependence of the
half-off-shell NN amplitudes $R^{++}(p',p)$
were exact.

We note that we did not employ the effective range approximation
for the final state scattering amplitude as it was used in Ref. 
\cite{KR66}; it was found in Ref. \cite {MS91} to artificially increase 
the cross section.  
Since the approximation  (\ref{mmat})  
allows for a very good description of the energy dependence
of the data, it can be used to extract the value of
the threshold amplitude ${\cal M}$. A fit to the data of  
Ref. \cite{Meyer} yields a value 
\begin{equation}
 | {\cal M}_{exp} | = 30.5 \times 10^{-7} {\rm MeV}^{-3} \, ,  
\label{aexp}
\end{equation}
which is to be compared with our calculations summarized in Table \ref{T1}.
A similar fit employing the unmodified phase space density with
an unrealistic energy dependence of the cross section leads to a much
smaller value $| {\cal M}_{exp} | = 5.0 \times 10^{-7}$\  MeV$^{-3}$.
  
\section{Results and conclusion}

Table \ref{T1} shows in detail the contributions of diagrams (a)-(c) for 
several
models that differ in their off-shell parameter $\nu$.
The best fit to the $NN$\ data was obtained \cite{StG} by models with
$\nu$\ in the range $1.5 - 2.0$. From the models listed in the table,
the one with $\nu = 1.6 $ yields the triton binding energy closest
to its experimental value \cite{StG}.  
 
The amplitudes are, in general, complex numbers, but we can subtract
the two contributions from diagram (c) which correspond to
having both intermediate nucleons on-shell either before or after
pion is emitted, and add them to the amplitudes of diagrams (a) and
(b), respectively. Then we arrive at new partial amplitudes 
that can be written  
as real numbers (which are given in Table \ref{T1}) times a 
common phase factor. However, one has to be
cautious when comparing to other calculations that present 
results for the original diagrams. At threshold, the final state
phase shift is zero and in terms of the real numbers $A$, $B$,  and $C$\
the actual partial amplitudes are given as
\begin{eqnarray}
  a^{\pm} &=& A^{\pm}/\cos \delta_3   \, \nonumber\\
  b^- &=& \exp ( i \delta_3 )\, B^-   \, \nonumber\\
  c^{\pm,+} &=& \exp ( i \delta_3 )\, ( C^{\pm,+} + 
   i A^{\pm} \tan \delta_3 ) \, \nonumber\\
 c^{\pm,-} &=&  \exp ( i \delta_3 )\, C^{\pm,-} \, .
 \end{eqnarray}

The upper part of the table presents results of our covariant 
calculations. Diagram (c) dominates at threshold.
Our results show a moderate dependence on the parameter $\nu$
and we observe that the amplitudes for the models with 
nonzero $\nu$\  are somewhat enhanced.
The largest value, at $\nu = 1$, is
about 20 \% larger than the smallest at  $\nu = 0$.
This translates into a more than 40 \% variation in the
total cross section. Although in itself it represents a
sizeable effect, it is at this point not clear how much of
it will survive in more complete calculations that would
include other, model-dependent reaction mechanisms (such as
that of diagram (d)). 
 
 Our total result is 2-3 times smaller than the fitted 
amplitude (\ref{aexp}) for all models. The relativistic
$NN$ potentials \cite{GVH} contain form factors for
off-shell nucleons (sometimes called ``sideways'' form
factors). For consistency we include them also in the
pion emission vertex. They decrease the results by up to
10  \% (see Table \ref{T1}). For reference and for comparison
with perturbative results  the total results without these form
factors are also given (line `` no N ffs ''). The nucleon form factors
are not included in our perturbative transition operators.
It is difficult to take them into account consistently in $v/c$-
expansion approaches, and the usual $NN$ potentials and pion
emission operators never consider them. 

The last three lines of Table \ref{T1} list the usual perturbative results
obtained with the same nucleon wave functions and with the
meson parameters and meson-nucleon form factors consistent with
the $NN$ interaction employed. The first line gives
the perturbative IA, to be compared with 
the subtotal ${\cal M }^+ = A^+ + C^{++}$\  
of covariant results. The perturbative result is somewhat
larger, the difference between these numbers is due to the
$v/c$-decomposition in the one-nucleon operator at threshold
(in the covariant description it contains a factor $1/E$\ compared to 
$1/m$ in the perturbative one) and to the presence of the off-shell 
nucleon form factors in the relativistic calculations.  

The perturbative contribution of the Z-diagrams given in the
second last line is to be compared to the sum of all terms
with negative energy nucleon(s) in intermediate states:
${\cal M }^- =A^- + B^- + C^{+-} + C^{-+} + C^{--}$. 
We see that the perturbative
result is more than 3 times as large as the covariant one. 
Notice also that the perturbative results would reproduce very well 
the experimental amplitude (\ref{aexp}), though one should recall
that Coulomb effects would suppress the cross section at threshold
by about 40 \%.

We do not consider here the model-dependent pion-production
mechanisms, e.g., the pion-rescattering, and the $\rho-\omega-\pi$\ diagram
introduced in Ref. \cite{Kol96b}.  They can 
easily contribute by the same amount as the model independent nucleon
Born diagrams considered so far. However, care should be taken to include
them in a way consistent with the dynamics of the $NN$-interaction.  
 
In conclusion, the sensitive cross section of $\pi^0$\ 
production seems  to be an
ideal place to look for effects of relativistic dynamics. Our exploratory
studies reported in this letter indeed lead to an assessment of the importance
of ``Z-diagrams'' different from the traditional nonrelativistic ones.
Our main conclusion is that such diagrams, when included in a
non-perturbative way, contribute differently than  their
non-relativistic limits taken perturbatively. A similar effect was 
reported in Ref. \cite{Tjon} for the elastic electron scattering on
deuteron, but there it becomes noticeable at a momentum transfer $\sim 1$\
GeV.  Our present calculations in relativistic impulse
approximation  underestimate the data.
In future calculations we plan to go beyond the threshold approximations
and include pion rescattering contributions. 

\acknowledgments

J.A.\ and F.G.\ thank the Lisbon CFNUL theory group, 
and A.S.\ and M.T.P.\ thank
the Jefferson Lab theory group, for the kind hospitality they received
during their mutual visits. 
We wish to thank S.\ Coon, Ch. Hanhart, D.O.\ Riska, and
U. van Kolck for helpful discussions on the subject.
This work was supported by
the DOE under contract number DE-FG05-88ER40435, by PRAXIS under
contract number praxis/2/2.1/FIS/223/94, by CERN under contract number
CERN/P/FIS/1101/96, and by JNICT under contract number BCC/4394/94.

%
\begin{figure}
\vbox{
\centerline{\epsfig{file=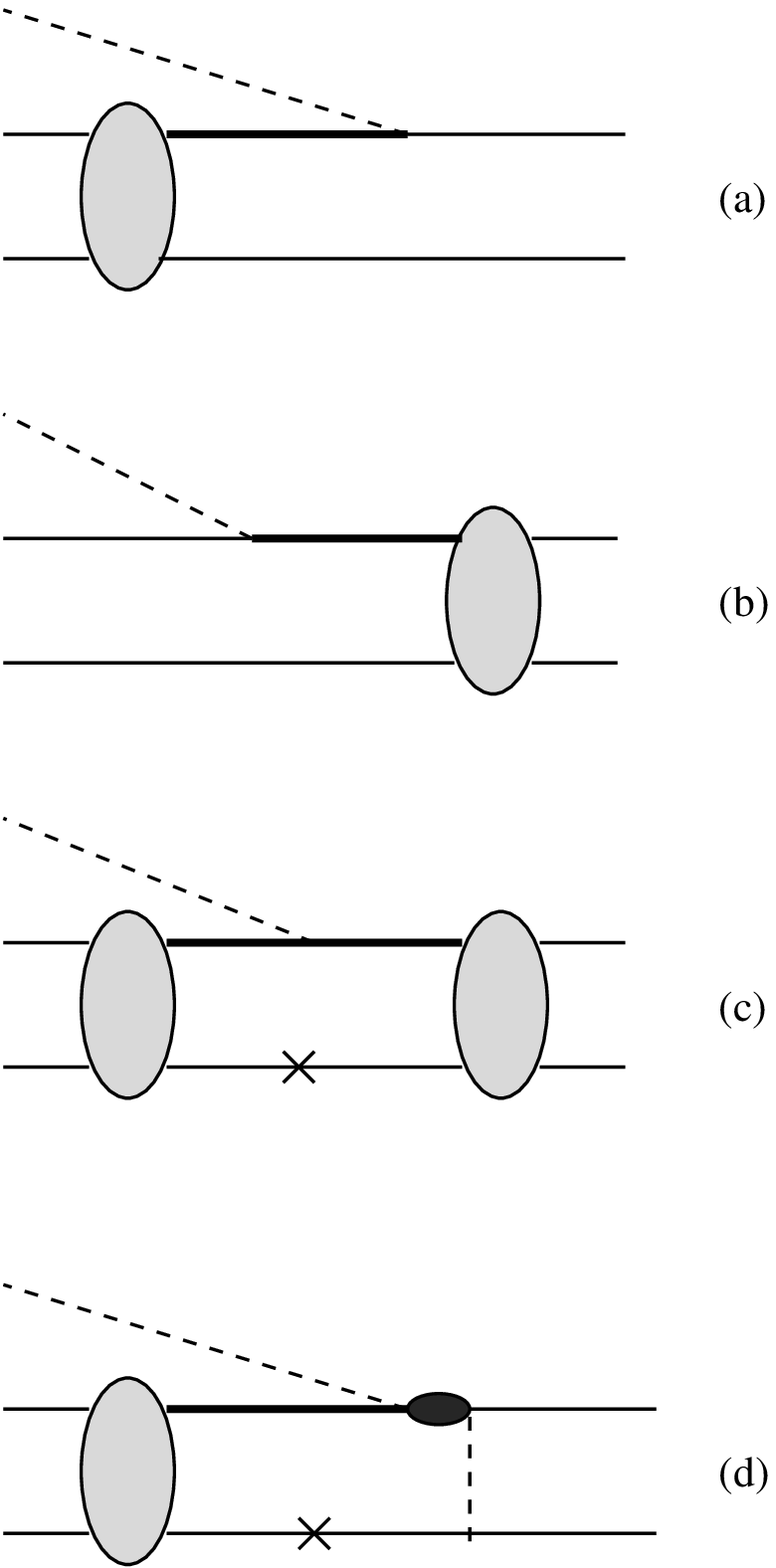,height=3.5in}}
\vspace*{5mm}
\caption{ The relativistic impulse approximation is defined
by diagrams (a), (b), and (c). The shaded areas are $NN$ scattering
amplitudes. Diagram (d) is an example of the model dependent
diagrams not included here. The thick lines represent off-shell nucleons,
they can propagate with positive and negative energy.}
\label{fig1}
}
\end{figure}
%
%
\begin{figure}
\vbox{
\centerline{\epsfig{file=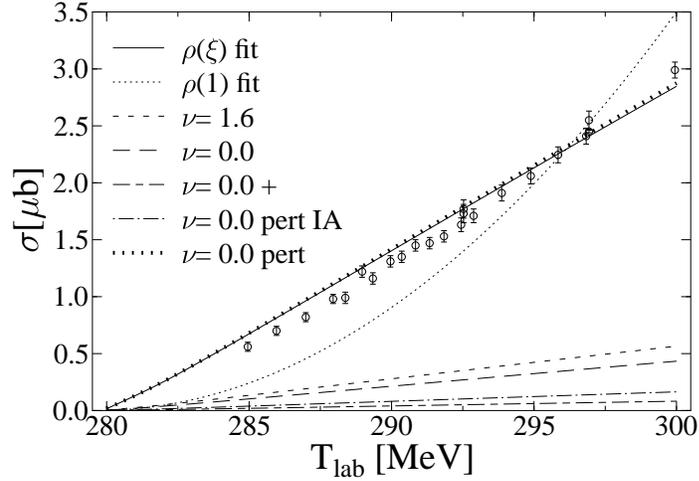,height=3.0in}}
\caption[99]{Total cross section vs lab energy of incoming proton. 
The Indiana data \cite{Meyer} are shown with the best fits using  
unmodified $\rho(1)$\ and modified $\rho(\xi)$\ phase space 
densities, respectively. The curves labeled  $\nu = 1.6$\
and  $\nu = 0.0$\ are the full results of the covariant calculations.
For the model with $\nu = 0.0$\ we also show  the cross section calculated 
with $\cal{M}^+$\ only, labeled `` $\nu = 0.0 \, +$'',
to be compared with the perturbative IA ``$\nu = 0.0$, pert IA''.  
The full perturbative prediction  ``$\nu = 0.0$, pert'' virtually
coincides with the  $\rho(\xi)$\ fit. }
\label{fig2}
}
\end{figure}
%
\begin{figure}
\vbox{
\centerline{\epsfig{file=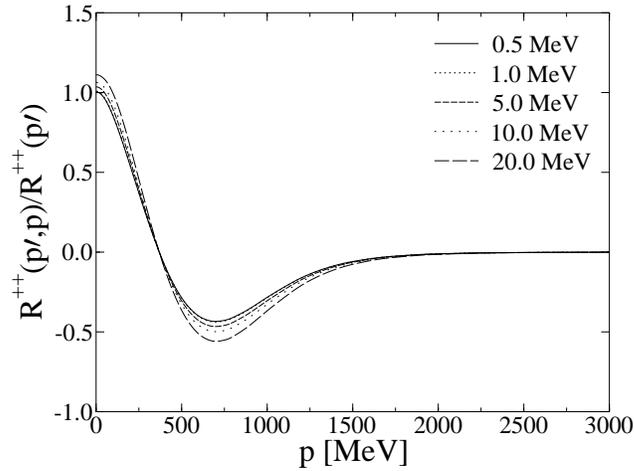,height=3.0in}}
\caption{Ratio of half off-shell and on-shell amplitudes of the model 
with $\nu = 1.6$\  for various two-nucleon lab energies (in MeV) as a
function of the off-shell momentum p. }
\label{fig3}
}
\end{figure}
%
%
%
\begin{table}[h]
\setdec 000.0000
\caption{Amplitudes  
of diagrams (a), (b), (c) for several $NN$-interaction models. For
diagrams (a)  and  (c) the amplitude is also separated into
contributions due to positive and negative energy nucleons
in the intermediate state(s). To diagram (b)  at threshold
only intermediate negative energy state contributes. 
${\cal M }^+ = A^+ + C^{++} $, ${\cal M }^- = A^- + B^- +
C^{+-} + C^{-+} + C^{--}$. The correspondence 
between the real numbers $A, B$,
and $C$\ and the complex amplitudes of diagrams (a), (b), and (c), is
explained in the text. The label ``no N ffs'' refers to the
total covariant results with the off-shell nucleon form factors
switched off. Perturbative results are denoted by ``Pert.''.
All amplitudes are given in units of $10^{-7}\,  {\rm MeV}^{-3}$.}
\vspace*{5mm} 
\begin{tabular}{l r r r r}
 Diagram  & $\nu = 0$ & $\nu = 1.0 $ & $ \nu = 1.6$ & $\nu = 2.0$ \\
\tableline
  $A^+ $           & $ 1.11$    & $-0.70$ & $-1.08$ & $-1.28$ \\
  $A^- $           & $-1.44$    & $-0.87$ & $-0.04$ & $ 0.68$ \\ 
  $A  $            & $-0.33$    & $-1.58$ & $-1.12$ & $-0.60$ \\
\tableline
  $B= B^-$         & $-0.47$    & $-0.55$ & $-0.55$ & $-0.54$ \\
\tableline
  $C^{++} $        & $-6.28$    & $-6.06$ & $-5.73$ & $-5.62$ \\     
  $C^{+-} $        & $-5.42$    & $-6.50$ & $-6.41$ & $-6.28$ \\   
  $C^{-+} $        & $ 0.56$    & $ 0.42$ & $ 0.19$ & $ 0.02$ \\   
  $C^{--} $        & $ 0.06$    & $ 0.06$ & $ 0.04$ & $ 0.01$ \\     
  $C      $        & $-11.1$    & $-12.1$ & $-11.9$ & $-11.9$ \\ 
\tableline
 $ {\cal M }^+ $   & $-5.17 $   & $ -6.77  $ & $ -6.81 $ & $ -6.90$ \\
 $ {\cal M }^- $   & $-6.71 $   & $ -7.44  $ & $ -6.77 $ & $ -6.11$ \\
 Total             & $-11.88$   & $-14.21  $ & $-13.57 $ & $-13.00$ \\
 no N ffs          & $-12.64$   & $-14.70  $ & $-13.91 $ & $-13.30$\\
\tableline
 Pert. IA          & $-7.31 $   & $-8.72 $ & $-8.50 $ & $-8.48 $ \\
 Pert. Z-diag      & $-23.3 $   & $-24.6 $ & $-24.8 $ & $-23.7 $ \\      
 Pert. total       & $-30.6 $   & $-33.3 $ & $-33.3 $ & $-32.2 $ \\  
\end{tabular}
\label{T1}
\end{table}

\end{document}